\documentclass[12pt]{article} 

\usepackage{comment}
\usepackage{latexsym}
\usepackage{amssymb,amsfonts}

\usepackage[utf8]{inputenc}
\usepackage[T1]{fontenc} 
\usepackage[english]{babel}           

\usepackage[pdftex]{graphicx}
\usepackage{epsfig}
\usepackage{graphicx}
\usepackage{comment}
\usepackage{latexsym}
\usepackage{hyperref}
\usepackage{amsmath}
\usepackage{mathrsfs}
\usepackage{calrsfs} 
\DeclareMathAlphabet{\pazocal}{OMS}{zplm}{m}{n} 
\usepackage[usenames, dvipsnames]{color}
\usepackage{amsbsy}
\usepackage{amssymb}
\usepackage{amsthm}
\usepackage{amsfonts}
\usepackage{cite}
\usepackage{enumitem}
\usepackage{xcolor}
\usepackage{color}
\usepackage{mathtools}
\usepackage{caption} 
\usepackage{subcaption} 
\usepackage{esvect}
\usepackage{cancel}

\usepackage{diagbox}

\usepackage{tikz}

\usepackage{tcolorbox}

\newcommand{\be}[0]{\begin{equation}}
\newcommand{\ee}[0]{\end{equation}}

\renewcommand{\thefootnote}{\fnsymbol{footnote}}

\newcommand{\Z}{\mathbb{Z}}
\renewcommand{\natural}{\mathbb{N}}

\renewcommand{\O}{{\pazocal O}}
\renewcommand{\Re}{{\rm Re}\,}
\renewcommand{\Im}{{\rm Im}\,}

\newcommand{\tr}{\textrm{tr}\,}

\renewcommand{\d}{\text{d}}
\newcommand{\e}{e}

\newcommand{\where}{\mbox{where}}

\renewcommand{\and}{\mbox{and}}

\newcommand{\esp}{\phantom{\!\!\overset{\displaystyle |}{.}}}
\newcommand{\espD}{\phantom{\!\!\underset{\scriptstyle |}{\cdot}}}

\newcommand{\bm}{\boldmath} 

\newcommand{\red}{\color{red}}

\newcommand{\cM}{\mathcal{M}}
\newcommand{\cN}{\mathcal{N}}
\newcommand{\cC}{\mathcal{C}}
\newcommand{\cA}{\mathcal{A}}

\newcommand{\N}{{\pazocal N}}
\renewcommand{\L}{{\pazocal L}}

\newcommand{\C}{{\pazocal C}}
\newcommand{\cR}{{\pazocal R}}

\newcommand{\m}{{\vec m}}
\newcommand{\vell}{{\vec \ell}}

\catcode`\@=11
\def\marginnote#1{}
\newcount\hour
\newcount\minute
\newtoks\amorpm
\hour=\time\divide\hour by60 \minute=\time{\multiply\hour by60
\global\advance\minute by-\hour}
\edef\standardtime{{\ifnum\hour<12 \global\amorpm={am}%
        \else\global\amorpm={pm}\advance\hour by-12 \fi
        \ifnum\hour=0 \hour=12 \fi
        \number\hour:\ifnum\minute<10 0\fi\number\minute\the\amorpm}}
\edef\militarytime{\number\hour:\ifnum\minute<10 0\fi\number\minute}
\def\draftlabel#1{{\@bsphack\if@filesw {\let\thepage\relax
   \xdef\@gtempa{\write\@auxout{\string
      \newlabel{#1}{{\@currentlabel}{\thepage}}}}}\@gtempa
   \if@nobreak \ifvmode\nobreak\fi\fi\fi\@esphack}
        \gdef\@eqnlabel{#1}}
\def\@eqnlabel{}
\def\@vacuum{}
\def\draftmarginnote#1{\marginpar{\raggedright\scriptsize\tt#1}}
\def\draft{\oddsidemargin -.2truein
        \def\@oddfoot{\sl preliminary draft \hfil
        \rm\thepage\hfil\sl\today\quad\militarytime}
        \let\@evenfoot\@oddfoot \overfullrule 3pt
        \let\label=\draftlabel
        \let\marginnote=\draftmarginnote
   \def\@eqnnum{(\theequation)\rlap{\kern\marginparsep\tt\@eqnlabel}%
\global\let\@eqnlabel\@vacuum}  }
\def\thebibliography#1{
\vskip 0.5cm \centerline{\bf \Large References}
\list{
[\arabic{enumi}]}{\settowidth\labelwidth{[#1]}
\leftmargin\labelwidth
\advance\leftmargin\labelsep
\usecounter{enumi}}
\def\newblock{\hskip .11em plus .33em minus .07em}
\sloppy\clubpenalty4000\widowpenalty4000
\sfcode`\.=1000\relax}

\renewcommand{\theequation}{\arabic{section}.\arabic{equation}}
\renewcommand{\section}{\setcounter{equation}{0}\@startsection
{section}{1}{0mm}{-\baselineskip}{0.5\baselineskip} {\normalfont\Large\bfseries}}
\renewcommand{\subsection}{\@startsection
{subsection}{2}{0mm}{-\baselineskip}{0.5\baselineskip} {\normalfont\large\bfseries}}
\renewcommand{\subsubsection}{\@startsection
{subsubsection}{3}{0mm}{-\baselineskip}{0.5\baselineskip}
{\normalfont\normalsize\slshape}}

\begin{document}


\begin{titlepage}
\begin{flushright}
CPHT-RR102.122019, December   2019
\vspace{1cm}
\end{flushright}
\begin{centering}
{\bm\bf \Large INDUCED EINSTEIN GRAVITY \\ \vspace{.2cm} FROM INFINITE TOWERS OF STATES  
 }

\vspace{5mm}

 {\bf A. Kehagias$^{1}$, H. Partouche$^2$ and B. de Vaulchier$^2$}

 \vspace{3mm}

$^1$ Physics Division, National Technical University of Athens,\\
15780 Zografou Campus, Athens, Greece\\
{\em  kehagias@central.ntua.gr}

$^2$  {CPHT, CNRS, Ecole polytechnique, IP Paris, \\F-91128 Palaiseau, France\\
{\em herve.partouche@polytechnique.edu, \\ balthazar.devaulchier@polytechnique.edu}}

\end{centering}
\vspace{0.7cm}
$~$\\
\centerline{\bf\Large Abstract}\\

\begin{quote}

We  consider four-dimensional quadratic gravity coupled to infinite towers of free massive scalar fields, Weyl fermions and vector bosons.  
We find that for specific numbers of  towers, finite cosmological and Newton constants are induced in the 1-loop effective action. 
This is derived both in Adler's approach and by using the heat kernel method, which yield  identical results. If the infinite number of massive states may be regarded as  Kaluza--Klein modes arising from fields in higher dimensions, there are no Kaluza--Klein states associated with the four-dimensional graviton.  Hence gravity is intrinsically four-dimensional.

\end{quote}

\end{titlepage}
\newpage
\setcounter{footnote}{0}
\renewcommand{\thefootnote}{\arabic{footnote}}
 \setlength{\baselineskip}{.7cm} \setlength{\parskip}{.2cm}

\setcounter{section}{0}


\section{Introduction}

Induced gravity is an old proposal \cite{zah,Adler1,Adler2,Adler3,Adler4,Adler5,zee1,zee2,zee3,Fro, Don,Visser} according to which gravity is not fundamental but is rather induced by quantum effects from the matter content of the universe. In other words, the gravitational dynamics and the associated curvature of spacetime are emerging phenomena,  a mean field approximation of the underlying microscopic degrees of freedom. In this approach, Einstein gravity is similar to  fluid dynamics which is a macroscopic approximation of Bose--Einstein condensates. 

In the induced gravity framework, one assumes a prescribed but not dynamical background on which matter fields are living. The latter are scalars, spinors and vectors coupled to the spacetime metric, which  is clearly not dynamical as it appears with no derivatives. For spinors in particular, one also regards the spin connection of the background  non dynamical as well. 
However, although the classical action does not contain derivatives of the metric, this is not true at the quantum level. Indeed, the  cosmological and Einstein-Hilbert terms of General Relativity are induced by loop corrections. This is the induced gravity proposal: The dynamics of gravity emerges from the quantum fluctuations of matter fields. However, a minimum requirement in this approach is that the induced Einstein term is finite. Otherwise, a tree-level Einstein counterterm would be needed to absorb the infinities, which would spoil the whole picture from the outset.  

In the present work, we first consider the above approach within the formalism of Adler~\cite{Adler1,Adler2,Adler3,Adler4,Adler5}, where the effective Einstein action parameters arise by integrating out massive matter fields. The expressions of the cosmological and Newton constants  are given by\footnote{Our choice of spacetime signature is $(-,+,+,+)$.}
\be\label{adler_form}
\begin{aligned}
\frac{1}{8\pi}\frac{\Lambda_{\rm ind}}{G_{\rm ind}} =& -{1\over 4}\, \langle T(0)\rangle\, , \\
\frac{1}{8\pi G_{\rm ind}} =& -\frac{i}{48}\int\d^4x\; x^2\, \langle\widetilde{T}(x)\widetilde{T}(0)\rangle\, ,
\end{aligned}
\ee
where $T$ is the trace of the stress-energy tensor of the matter fields. Moreover,  $\widetilde{\O}(x)=\O(x)-\langle \O(x)\rangle$ is the variation of any observable $\O$ with respect to its vacuum expectation value, and all correlation functions are computed in Minkowski spacetime.  We show that a particular choice of matter content yields finite $\Lambda_{\rm ind}$ and $G_{\rm ind}$. The spectrum is naturally interpreted from a higher dimensional  perspective. We assume a $(4+n)$-dimensional spacetime of the form $M^4\times S^1\times\cdots \times S^1$, where the $n$  circles have radii $R_i$, $i\in\{1,\dots,n\}$, and $M^4$ is a four-dimensional curved spacetime.  The  matter fields living in $M^4$ are the Kaluza-Klein (KK) modes of free fields in $4+n$ dimensions. Notice that Adler's approach has also been employed in the DGP model~\cite{DGP}, where a localized four-dimensional Einstein-Hilbert action is induced on a brane sitting at a point of a fifth infinite dimension.
However, a crucial difference in our setup is that gravity is only four-dimensional, which means that the graviton propagates in four dimensions and cannot see the extra directions. This is at odds with the usual brane-world scenario, where the opposite happens, namely that the graviton lives in higher dimension, while the matter fields are localized on a four-dimensional subspace. In the latter case, the weakness of gravity with respect to the other interactions is a consequence of the presence of extra space available for the graviton propagation.

However, in order to justify that  the induced gravity action is to be extremized with respect to the metric, it may be unavoidable to treat the graviton as a quantum field. Another concern is that higher derivative terms of the metric are also induced by the loops of matter fields. This is more easily seen by using heat kernel methods~\cite{zah,dewitt, Gilkey, Duff1,Duff2,BD, Toms, Avramidi2,Avramidi3, PT} to derive the 1-PI effective action. All of these terms turn out to be finite, except those with four derivatives. Hence, counterterms of the same form should be introduced at tree level. If the metric is to be quantized, then the general picture falls within the framework of the so-called Quadratic Gravity~\cite{Stelle1,Stelle2,AG,Salvio,Shapiro}. Such theories are renormalizable, even in presence of matter. In that case, renormalized cosmological and Newton constants are generated but not predictable, as follows from their running under the renormalization group flow. However, for infinite towers of matters states, we find that the Einstein action parameters  $\Lambda_{\rm ind}$ and $G_{\rm ind}$ remain predictable. This is achieved at the semiclassical level for the gravitational degrees of freedom and still holds at the 1-loop level. Whether this statement continues to be true at higher loop order of the gravity sector is a question that is beyond the scope of the present work. 

In Sect.~\ref{IEA}, treating the metric as a classical background, we derive in the context of Adler's formalism the finite cosmological and Newton constants that are induced by integrating out infinite towers of free real scalars, Weyl fermions and vector bosons.  In Sect.~\ref{aux}, we show that including non-dynamical or auxiliary fields in the classical  theory does not alter the final result for the effective Einstein action parameters. In Sect.~\ref{sak}, we argue that the gravitational degrees of freedom should be quantized in the context of Quadratic Gravity. In presence of towers of matter fields, finiteness and predictability of all gravitational terms is ensured at the 1-loop level, except for the Ricci square and Weyl square terms, which are running. In Sect.~\ref{conc}, we summarize our results and sketch how they may accommodate the Standard Model. However, more work for establishing whether this can be achieved is required. 


\section{Induced Einstein action}
\label{IEA}

In order to derive induced cosmological and gravitational constants $\Lambda_{\rm ind}$ and $G_{\rm ind}$ in four spacetime dimensions, we consider quantum free fields living in  a given $4+n$ dimensional background manifold, where  {\em  $n\ge 1$ space-like directions are circles of constant radii $R_i$, and the four-dimensional Lorentzian metric depends on the four-dimensional coordinates only,} $g_{\mu\nu}(x^\lambda)$. The higher dimensional origin of the degrees of freedom translates into infinite towers of Kaluza-Klein modes. In the following, we apply Adler's results~\cite{Adler1,Adler2,Adler3,Adler4,Adler5} given in Eq.~\eqref{adler_form} to towers of real scalar fields, Weyl fermions and vector fields. It is by choosing suitably the initial spectrum in $4+n$ dimensions that a finite four-dimensional theory for gravity  is  obtained at the two-derivative level. 


\subsection{Real scalar field}

In a four-dimensional spacetime with metric $g_{\mu\nu}$ of signature $(-,+,+,+)$, the action of the KK states arising from a real scalar free field in $4+n$ dimensions is
\be
S_\phi=-\int \d^4x\sqrt{-g}\,  \frac{1}{2}\sum_\m \Big[g^{\mu\nu}\partial_\mu\phi_\m\partial_\nu\phi_{\vec m} + M_\m^2\phi_\m^2 \Big],
\label{scalar}
\ee
where $\m\equiv (m_4,\dots,m_{3+n})\in \Z^n$ labels the mode $\phi_\m$ of squared mass 
\be
M_{\m}(\vec Q)^2 = \sum_{i=4}^{3+n}\left(\frac{m_i + Q_i}{R_i}\right)^2\, .
\label{mm}
\ee
In the above formula, we include a shift of the momenta by a non-trivial real  vector $\vec Q$, so that  none of the KK modes is massless. In practice, this means that we impose $\vec Q \notin \Z^n$. In that case, all states can be integrated out and a meaningful effective theory is obtained. Note that such a shift is allowed once a pair of real scalar fields $\phi$, $\tilde \phi$ in higher dimensions is combined into a complex scalar $\Phi\equiv (\phi+i\tilde \phi)/\sqrt{2}$, and that non-trivial boundary conditions are imposed at least along one compact direction. To be specific, we have 
\begin{equation}
\Phi(x^\mu,\vec x)= {1\over \sqrt{\prod_i 2\pi R_i}}\sum_{\vec m}\Phi_{\vec m}(x^\mu)\,  \e^{i\sum_j \!{m_j+ Q_j \over R_j}x^j}\, ,
\end{equation}
where $\Phi_{\vec m}\equiv (\phi_{\vec m}+i\tilde \phi_{\vec m})/\sqrt{2}$ and $\vec Q$ is a global $U(1)^n$ charge vector. 
The trace of the stress-energy tensor reads 
\be
g^{\mu\nu}T_{\mu\nu}^\phi\equiv {-2\over \sqrt{-g}}\, g^{\mu\nu}{\delta S_\phi\over \delta g^{\mu\nu}}=-\sum_\m \Big[g^{\mu\nu}\partial_\mu\phi_\m\partial_\nu\phi_\m +2 M_\m^2\phi_\m^2 \Big].
\ee
What is required for deriving the induced gravity action is the above expression in Minkowski spacetime ($g_{\mu\nu}=\eta_{\mu\nu}$), 
\be
T_\phi(x)= -\sum_\m \Big[\partial_\mu\phi_\m\partial^\mu\phi_\m + 2M_\m^2\phi_\m^2 \Big],
\ee
as well as the two-point functions  in flat space
\be
\langle \phi_\m(x)\phi_{\m'}(y)\rangle = \delta_{\m,\m'}\, \Delta_\m(x-y)\, , \quad \where \quad \Delta_\m(x-y) = -i\int\frac{\d^4k}{(2\pi)^4}\, \frac{\e^{ik\cdot (x-y)}}{k^2+M_\m^2-i\varepsilon}\, .
\label{2pt}
\ee

The contribution to the cosmological constant arising by integrating out the scalars $\phi_\m$ is proportional to the vacuum expectation value of $T_\phi$, which is\footnote{We have $-i(2\pi)^4\delta^{(4)}(0)= (2\pi)^4\delta^{(4)}(0_{\rm E})=V_4$, where $0_{\rm E}$ is the origin of the four-dimensional Euclidean spacetime of infinite volume $V_4$. } 
\be
\langle T_\phi(0)\rangle =\sum_\m\left[i\delta^{(4)}(0) -M_\m^2\, \Delta_\m (0)\right] .
\label{tphi}
\ee
Note that this expression is only formal for two reasons. On the one hand, the Dirac distribution at $x=0$, which arises for each mode $\m$, yields an infinite constant. However, the latter being mass- or radii-independent, it is free of any physical content and, as will be seen in Sect.~\ref{aux}, can be removed by adjusting the content of non-dynamical (for example auxiliary) fields in  the full theory. 
On the other hand, each Feynman propagator $\Delta_\m$ at $x=0$ leads to an UV quadratic divergence. As detailed in the Appendix, the latter can be dealt with by adopting a prescription inspired by string theory compactified on tori:
\begin{itemize}
\item We first apply a Wick rotation and switch to first quantized formalism by introducing a Schwinger parameter $t$ for the propagator. The key point is to perform the Schwinger integral in last. 
\item The Gaussian integral over the Euclidean momentum $k_{\rm E}$ is done first.
\item A Poisson summation is applied on the discrete KK sum over $\m$. 
\item In this form, all but one term (associated with the UV divergence) of the discrete sum can be integrated term by term over $t$.
\end{itemize}
In the end, the contribution to the induced cosmological constant given in Eq.~(\ref{adler_form}) and arising from the KK tower of scalar fields takes the form
\be
\left.\frac{1}{8\pi}\frac{\Lambda_{\rm ind}}{G_{\rm ind}}\right|_\phi = -{i\over 4}\sum_\m \delta^{(4)}(0) +I_0({\vec Q})\, ,
\label{lphi}
\ee
where we denote
\be
I_0({\vec Q}) = - \frac{\Gamma\big(2+\frac{n}{2}\big)}{32\pi^{6+{n\over2}}}\,\Big(\prod_{i=4}^{3+n} R_i\Big)\sum_{\vell}\frac{\e^{2i\pi \vec Q\cdot\vell}}{\big(\sum_j\ell_j^2R_j^2\big)^{2+{n\over2}}}\, ,
\ee
and where $\vell\equiv (\ell_4,\dots,\ell_{3+n})\in \Z^n$. As mentioned above, the UV divergence manifests itself as the term $\vell=\vec 0$. As will be shown in Sect.~\ref{smod}, it is the interplay between different KK towers of states  that will remove all ill-defined contributions $\vell=\vec 0$.

The correlator to be computed for deriving the induced gravitational constant is
\be
\label{TT_scal}
\langle\widetilde T_\phi(x)\widetilde T_\phi(0)\rangle =2\sum_\m \Big[ \partial_\mu\partial_\nu\Delta_\m(x)\,   \partial^\mu\partial^\nu\Delta_\m(x)+ 4M_\m^2\partial_\mu\Delta_\m(x) \,\partial^\mu\Delta_\m(x)+ 4M_\m^4\Delta_\m(x)^2 \Big].
\ee
From the definition~\eqref{adler_form}, we obtain
\be
\left.\frac{1}{8\pi G_{\rm ind}}\right|_\phi = I_1({\vec Q}) + I_2({\vec Q}) + I_3({\vec Q})\, ,
\ee
where we have defined
\be\label{integrals}
\begin{aligned}
I_1({\vec Q}) =&\;\,\; \frac{i}{24\, (2\pi)^8}\int\d^4x\,x^2\sum_{\vec m}\int \d^4p\,\d^4k\ \frac{(p\cdot k)^2\; \e^{i(p+k)\cdot x}}{\big(p^2+M_\m^2-i\varepsilon\big)\big(k^2+M_\m^2-i\varepsilon\big)}\, , \\
I_2({\vec Q}) =& -\frac{i}{6\, (2\pi)^8}\int\d^4x\,x^2\sum_{\vec m}M^2_{\vec m}\int \d^4p\,\d^4k\ \frac{ p\cdot k\; \e^{i(p+k)\cdot x}}{\big(p^2+M_\m^2-i\varepsilon\big)\big(k^2+M_\m^2-i\varepsilon\big)}\, , \\
I_3({\vec Q}) =&\, \;\;\;\;\frac{i}{6\, (2\pi)^8}\int\d^4x\,x^2\sum_{\vec m}M^4_{\vec m}\int \d^4p\,\d^4k\ \frac{\e^{i(p+k)\cdot x}}{\big(p^2+M_\m^2-i\varepsilon\big)\big(k^2+M_\m^2-i\varepsilon\big)}\, .
\end{aligned}
\ee
The above quantities, can be computed by following the steps listed above for the cosmological constant, up to the additional integration over the Euclidean spacetime variable $x_{\rm E}$. As seen in the Appendix, after integration over the Euclidean momenta $p_{\rm E}$, $k_{\rm E}$, the integral over $x_{\rm E}$ turns out to be Gaussian and can therefore be trivially computed at this stage. The final expressions take the forms
\be
I_1({\vec Q}) = -\frac{\Gamma\big(1+{n\over 2}\big)}{32 \pi^{4+{n\over 2}}}\, \Big(\prod_{i=4}^{3+n} R_i\Big)\sum_{\vell}\frac{\e^{2i\pi \vec Q\cdot\vell}}{\big(\sum_j\ell_j^2R_j^2\big)^{1+{n\over2}}}\, , \qquad
I_2({\vec Q}) = -\frac{4}{3}\, I_1({\vec Q})\, ,  \qquad 
I_3({\vec Q}) =0\, ,
\ee
where, as before, the divergent contributions $\vell =\vec 0$ will cancel out with those arising from other KK towers. Hence, we obtain
\be
\left.\frac{1}{8\pi G_{\rm ind}}\right|_\phi =-{1\over 3}\,I_1({\vec Q}) \, .
\label{gphi}
\ee


\subsection{Weyl fermion}

Let us proceed with the analogous contributions to the Einstein gravity action induced by integrated out massive  fermions.  
To be specific, we consider a KK tower of four-dimensional Weyl fermions $\psi_\m$ of masses $M_\m$ and that are free. Their action can be written as 
\be
\label{acfer}
S_\psi= \int\d^4x \sqrt{-g}\,  \frac{1}{2}\sum_\m \Big[i\nabla_\mu\bar \psi_\m \bar \sigma^\mu \psi_\m-i\bar \psi_\m \bar \sigma^\mu\nabla_\mu\psi_\m-M_\m \big(\psi_\m\psi_\m+\bar \psi_\m\bar \psi_\m\big)\Big] ,
\ee
where our conventions can be found in Ref.~\cite{wb}. For a Weyl fermion $\psi$, which contains two degrees of freedom,  we denote its complex conjugate by $\bar \psi$, and define $\psi\psi=\psi^\alpha\psi_\alpha$, $\bar\psi\bar \psi = \bar\psi_{\dot\alpha}\bar\psi^{\dot\alpha}$, where spinorial indices are raised and lowered with Lorentz invariant antisymmetric tensors  $\epsilon^{\alpha\beta}$, $\epsilon_{\alpha\beta}$ or $\epsilon^{\dot\alpha\dot\beta}$, $\epsilon_{\dot\alpha\dot\beta}$. Moreover, $\sigma^\mu_{\alpha\dot\alpha}$ and $\bar \sigma^{\mu\dot \alpha\alpha}$ are $2\times 2$ matrices that reduce in Minkowski spacetime to those given in Appendix~A of Ref.~\cite{wb}. 
The stress-energy tensor that appears in Eq.~\eqref{adler_form} is the symmetric Belifante tensor used in general relativity. Its trace, 
\be
g^{\mu\nu}T^\psi_{\mu\nu}\equiv {e_{c\mu}\over \sqrt{-g}}\,g^{\mu\nu}{\delta S_\psi\over \delta E^\nu_{\; \;c}}\, , 
\ee
is defined in terms of the vielbein $e^a_{\;\; \mu}$ and inverse vielbein $E^\mu_{\;\; a}$. In Minkowski spacetime ($e^a_{\;\; \mu}=\delta^a_\mu$), the above trace reads   
\be
T_\psi(x) =\sum_\m\left[\frac{3i}{2}\Big(\partial_\mu\bar\psi_\m\bar\sigma^\mu\psi_\m - \bar\psi_\m\bar\sigma^\mu\partial_\mu\psi_\m\Big) -2M_\m\Big(\psi_\m \psi_\m+\bar\psi_\m\bar\psi_\m\Big)\right] ,
\ee
while the two-point functions are 
\be
\begin{aligned}
\langle\psi_{\m\alpha}(x)\psi_{\m'}^\beta(y)\rangle =&\ \delta_{\m\m'}\,\delta_\alpha^\beta \,M_\m\, \Delta_\m(x-y)\, , \\
\langle\bar\psi_{\m}^{\dot\alpha}(x)\bar\psi_{\m'\dot\beta}(y)\rangle =&\ \delta_{\m\m'}\, \delta_{\dot\beta}^{\dot\alpha}\,  M_\m\, \Delta_\m(x-y)\, , \\
\langle\psi_{\m\alpha}(x)\bar\psi_{m'\dot\beta}(y)\rangle =&\ -i\delta_{\m\m'}\,\sigma_{\alpha\dot\beta}^\mu\, \partial_\mu\Delta_\m(x-y)\, . \\
\end{aligned}
\ee

The correlator involved in the cosmological term takes the following form,
\be
\langle T_\psi(0)\rangle =\sum_\m\left[-6i\delta^{(4)}(0) +2M_\m^2\, \Delta_\m (0)\right] ,
\label{tpsi}
\ee
which yields
\be
\left.\frac{1}{8\pi}\frac{\Lambda_{\rm ind}}{G_{\rm ind}}\right|_\psi = {3i\over 2}\sum_\m \delta^{(4)}(0) - 2I_0({\vec Q})\, .
\ee
Some remarks are in order:
\begin{itemize}
\item The physical contribution $-2I_0({\vec Q})$, which is mass- or radii-dependent,  is twice the opposite of that found for a KK tower of real scalar fields. This is consistent with the fact that the vacuum energies arising from quantum fluctuations of bosonic and fermionic degrees of freedoms are opposite.
\item On the contrary, the Dirac distribution terms do not respect this rule, suggesting again that they are unphysical. 
\item Actually,  $I_0({\vec Q})$ reproduces exactly the expression of the  1-loop Coleman--Weinberg vacuum energy associated with a KK tower of degrees of freedom (see Sect.~\ref{sak}).    
\end{itemize}

To derive the contribution to the induced gravitational constant, we consider the two-point function
\be
\label{TT_psi}
\begin{aligned}
\langle\widetilde T_\psi(x)\widetilde T_\psi(0)\rangle=\sum_\m\Big[ \!-9\big(\square\Delta_\m\big)^2 + 9\partial_\mu\Delta_\m\partial^\mu\square\Delta_\m &+ 39M_\m^2\Delta_\m\square\Delta_\m\\
&- 7M_\m^2\partial_\mu\Delta_\m\partial^\mu\Delta_\m- 32M_\m^4\Delta_\m^2  \Big].
\end{aligned}
\ee
The latter yields  
\be
\left.\frac{1}{8\pi G_{\rm ind}}\right|_\psi  = I_4({\vec Q})  + I_5({\vec Q}) + I_6({\vec Q}) - \frac{7}{8}\,I_2({\vec Q})- 4\,I_3({\vec Q})\, ,
\ee
which involves $I_2({\vec Q})$ and $I_3({\vec Q})$ computed before, as well as similar quantities defined as 
\be\label{integrals2}
\begin{aligned}
I_4({\vec Q}) =&-\frac{3i}{16\, (2\pi)^8}\int\d^4x\,x^2\sum_{\vec m}\int \d^4p\,\d^4k\ \frac{p^2\, k^2\; \e^{i(p+k)\cdot x}}{\big(p^2+M_\m^2-i\varepsilon\big)\big(k^2+M_\m^2-i\varepsilon\big)}\, , \\
I_5({\vec Q}) =& \;\;\;\;\frac{3i}{16\, (2\pi)^8}\int\d^4x\,x^2\sum_{\vec m}\int \d^4p\,\d^4k\ \frac{ p^2\, (p\cdot k)\; \e^{i(p+k)\cdot x}}{\big(p^2+M_\m^2-i\varepsilon\big)\big(k^2+M_\m^2-i\varepsilon\big)}\, , \\
I_6({\vec Q}) =&-\frac{13i}{16\, (2\pi)^8}\int\d^4x\,x^2\sum_{\vec m}M^2_{\vec m}\int \d^4p\,\d^4k\ \frac{p^2\; \e^{i(p+k)\cdot x}}{\big(p^2+M_\m^2-i\varepsilon\big)\big(k^2+M_\m^2-i\varepsilon\big)}\, .
\end{aligned}
\ee
Proceeding as before, the latter are found to be
\be
I_4({\vec Q}) = 0 \, , \qquad I_5({\vec Q}) =-{3\over 2}\, I_1({\vec Q})\, , \qquad I_6({\vec Q}) =0\, , 
\ee
which leads to
\be
\left.\frac{1}{8\pi G_{\rm ind}}\right|_\psi =-{1\over 3}\,I_1({\vec Q}) \, .
\ee


\subsection{Vector field}

The last contribution to the induced gravity action we consider is that arising by integrating out a KK tower of vector bosons $A_\m^\mu$ of masses $M_\m$. Defining the field strength $F_{\m \mu\nu}=\partial_\mu A_{\m\nu}-\partial_\nu A_{\m\mu}$, the action of the massive spin-1 fields is 
\be
\label{acvb}
S_{A}= -\int\d^4x \sqrt{-g}\sum_\m \Big[ \frac{1}{4}\, g^{\mu\rho}g^{\nu\sigma}F_{\m\mu\nu}F_{\m\rho\sigma}+{1\over 2}\, M_\m^2\, g^{\mu\nu}A_{\m\mu}A_{\m\nu}\Big],
\ee
from which the trace of the stress-energy tensor is found to be 
\be
g^{\mu\nu}T_{\mu\nu}^A\equiv {-2\over \sqrt{-g}}\, g^{\mu\nu}{\delta S_A\over \delta g^{\mu\nu}}=-\sum_\m M_\m^2\, g^{\mu\nu} A_{\m\mu}A_{\m\nu} \, .
\ee
In Minkowski space, the latter reduces to 
\be
T_A(x) =-\sum_\m M_\m^2\, A_\m^\mu A_{\m\mu} \, ,
\ee
while the Feynman propagator in unitary gauge is\be
\langle A_{\m\mu}(x)A_{\m'\nu}(y)\rangle = \delta_{\m\m'}\!\left(\eta_{\mu\nu}-\frac{\partial_\mu\partial_\nu}{M_\m^2}\right)\Delta_\m(x-y)\, . 
\ee

The vacuum expectation value of the trace reads
\be
\langle T_A(0)\rangle = \sum_\m\left[i\delta^{(4)}(0) -3M_\m^2\, \Delta_\m (0)\right]\!,
\ee
where the factor $3$ is the number of physical degrees of freedom of a massive vector boson in unitary gauge. Therefore, the contribution to the cosmological constant is
\be
\left.\frac{1}{8\pi}\frac{\Lambda_{\rm ind}}{G_{\rm ind}}\right|_A = -{i\over 4}\sum_\m \delta^{(4)}(0) +3I_0({\vec Q})\, .
\label{al}
\ee
Moreover, the two-point function of $\widetilde{T}_A$ is
\be
\langle\widetilde{T}_A(x)\widetilde{T}_A(0)\rangle=2\sum_\m\Big[ \big(\partial_\mu\partial_\nu\Delta_\m\big)^2  - 2M_\m^2\Delta_\m\partial^2\Delta_\m+ 4M_\m^4\Delta_\m^2 \Big],
\ee
which yields
\be
\begin{aligned}
\left.\frac{1}{8\pi G_{\rm ind}}\right|_A &= I_1({\vec Q}) - \frac{4}{39}\, I_6({\vec Q}) + I_3({\vec Q})\\
&=I_1({\vec Q})\, .
\end{aligned}
\label{gl}
\ee

The pair of ghosts associated with each KK vector boson of internal momentum $\m$ must also be taken into account. Because in unitary gauge their masses are 
\be
\xi M_\m^2=\sum_{i}\left({m_i + Q_i \over R_i/\sqrt{\xi}}\right)^2\, , \quad \where\quad \xi\to +\infty\, ,
\ee
their contributions to the induced Einstein action can be deduced from those of a tower of complex scalar fields, in the case of vanishing rescaled radii, $R_i/\sqrt{\xi}\to 0$. Defining $T_{\text{ghosts}}$ to be the trace of the stress-energy tensor of the KK towers of ghosts in Minkowski space, we find from  Eqs~\eqref{tphi} and~\eqref{TT_scal} that
\be
\begin{aligned}
\langle T_{\text{ghosts}}(0)\rangle &= -4\sum_\m i\delta^{(4)}(0)\, ,\\
\langle\widetilde T_{\text{ghosts}}(x)\widetilde T_{\text{ghosts}}(0)\rangle &= -16\sum_\m \big[M_\m^4\Delta_\m(x)^2\big]\!\big |_{R_i/\sqrt{\xi}} \, ,
\end{aligned}
\ee
where we have added overall minus signs arising from the  anticommuting nature of the fields. As a result, the contributions of the ghosts to the gravitational constants are 
\be
\begin{aligned}
\left.\frac{1}{8\pi}\frac{\Lambda_{\rm ind}}{G_{\rm ind}}\right|_{\text{ghosts}} =&\  i\sum_\m \delta^{(4)}(0) \, , \\
\left.\frac{1}{8\pi G_{\rm ind}}\right|_{\text{ghosts}} =&\ -2I_3(\vec{Q})\big |_{R_i/\sqrt{\xi}}  = 0\, .
\end{aligned}
\ee


\subsection{Cancellation of UV divergences}
\label{smod}

For each KK tower of spin 0, ${1\over 2}$ or 1 fields which are integrated out, the induced cosmological term contains an infinite contribution proportional to $i\sum_\m \delta^{(4)}(0)$ that will be treated in the next section. For the time being, we focus on 
the quadratic and quartic UV divergences, which  appear respectively in the expressions of the cosmological and Newtonian constants. In our prescription, which is based on  first quantized formalism, these pathologies show up via infinite contributions for $\vell=\vec 0$.  Our goal is to determine the spectrum in $4+n$ dimensions for which  these divergences cancel out. 

Let us consider $N_\phi$, $N_\psi$, $N_A$ KK towers of real scalars, Weyl fermions and vector bosons (accompanied with their pairs of ghosts), and their associated charges $\vec Q_u\notin \Z^n$, $u=1,\dots,N_\psi+N_\psi+N_A$. Once integrated out, the total constants appearing in the effective action are
\be
\begin{aligned}
\frac{1}{8\pi}\frac{\Lambda_{\rm ind}}{G_{\rm ind}}=&\sum_{u=1}^{N_\phi}I_0(\vec Q_u)-2\sum_{u=N_\phi+1}^{N_\phi+N_\psi} I_0(\vec Q_u)+3\sum_{u=N_\psi+N_\psi+1}^{N_\phi+N_\psi+N_A}I_0(\vec Q_u)\\
&-\big(N_\Phi-6 N_\Psi+N_A-4N_A\big){i\over 4}\sum_\m \delta^{(4)}(0)\, , 
\\ \frac{1}{8\pi G_{\rm ind}}=&-{1\over 3}\sum_{u=1}^{N_\phi}I_1(\vec Q_u)-{1\over 3}\sum_{u=N_\phi+1}^{N_\phi+N_\psi} I_1(\vec Q_u)+\sum_{u=N_\psi+N_\psi+1}^{N_\phi+N_\psi+N_A}I_1(\vec Q_u)\, .
\label{resul}
\end{aligned}
\ee
The key point is that all contributions $\vell=\vec 0$ appearing in the quantities $I_0(\vec Q_u)$ and $I_1(\vec Q_u)$ are charge independent. Hence, for these terms to cancel, is is necessary and sufficient to satisfy the conditions
\be
N_\phi-2N_\psi+3N_A=0\, , \qquad -{1\over 3}\, N_\phi-{1\over 3}\, N_\psi+N_A=0\, ,
\ee
whose solutions are 
\be
(N_\phi,N_\psi,N_A)=(1,2,1) N, \quad N\in \natural\, .
\label{spec}
\ee
Notice that this result is {\em irrespective} to the choice of $N_\phi+N_\psi+N_A=4N$ charge vectors $\vec Q_u$. Therefore, there is no symmetry condition to assume between the KK towers of free fields for the effective Einstein Lagrangian to be finite.  However, as mentioned at the beginning of Sect.~\ref{IEA},  for the momentum shift $\vec Q$ to be interpreted as a global $U(1)^n$ charge vector in $4+n$ dimensions, we may impose $N$ to be even and have only pairs of KK towers with identical charges. 

To illustrate the above results, let us present some simple  cases of charge configurations. The simplest example we may consider is that of a universal value of the charges, namely $\vec Q_u\equiv \vec Q$, $u=1,\dots, 4N$. In this case, at each mass level $M_\m$, there is a vector boson, a real scalar and two Weyl fermions, which correspond to the field content of a massive vector multiplet of an exact $\N_4=1$ supersymmetry. However, the conditions for cancelling the $\vell=\vec 0$ terms imply all other contributions $\vell \neq \vec 0$ to vanish as well, so that no gravitational action (up to the Dirac distribution terms) is induced in this case,
\be
\begin{aligned}
\frac{1}{8\pi}\frac{\Lambda_{\rm ind}}{G_{\rm ind}}&=N\big(1-2\times 2+3\big)I_0(\vec Q)+N{7i\over 2}\sum_\m \delta^{(4)}(0)=0+N{7i\over 2}\sum_\m \delta^{(4)}(0)\, , 
\\ \frac{1}{8\pi G_{\rm ind}}&=N\big(\!-{1\over 3}-{1\over 3}\times 2+1\big)I_1(\vec Q)=0\, .
\end{aligned}
\ee

Non-trivial examples of effective Einstein actions correspond to the cases of spontaneous breaking of the $\N_4=1$ supersymmetry. They are realized by a Scherk--Schwarz mechanism~\cite{SS1,SS2} which consists in lifting the degeneracy between the KK masses of the bosons and fermions.  The simplest choice of charge vectors assumes a universal mass shift $\vec Q$, plus an additional half-integer shift $\delta \vec Q$ for all fermions,
\be
\mbox{Bosons : }\; \vec Q_u=\vec Q\, ,\qquad \mbox{Fermions : }\; \vec Q_u=\vec Q+\delta \vec Q\, , \;\;\delta Q_i\in\Big\{0,{1\over 2}\Big\}\, , i=1,\dots,n, \;\; \delta \vec Q\neq \vec 0\, .  
\ee
In this case, the sums over $\vell$ involve a projector $(1-\e^{2i\pi\delta\vec Q\cdot \vell})/2$ that eliminates an infinite number of contributions $\vell$, among which there are the dangerous terms $\vell=\vec 0$. The induced constants~(\ref{resul}) can then be explicitly written as
\be
\begin{aligned}
\frac{1}{8\pi}\frac{\Lambda_{\rm ind}}{G_{\rm ind}}&= -4N \, \frac{\Gamma\big(2+\frac{n}{2}\big)}{32\pi^{6+{n\over2}}}\,\Big(\prod_{i=4}^{3+n} R_i\Big)\sum_{\vec k}\frac{\e^{2i\pi \vec Q\cdot\vell}\, \big(1-(-1)^{2\delta \vec Q\cdot \vell}\big)}{\big(\sum_j \ell_j^2R_j^2\big)^{2+{n\over2}}}+N{7i\over 2}\sum_\m \delta^{(4)}(0)\, , \\
\frac{1}{8\pi G_{\rm ind}}&= N\, \frac{\Gamma\big(1+{n\over 2}\big)}{48 \pi^{4+{n\over 2}}}\, \Big(\prod_{i=4}^{3+n} R_i\Big)\sum_{\vell}\frac{\e^{2i\pi \vec Q\cdot\vell}\, \big(1-(-1)^{2\delta \vec Q\cdot \vell}\big)}{\big(\sum_j\ell_j^2R_j^2\big)^{1+{n\over2}}}\, .
\end{aligned}
\ee
Specializing for instance to the case of a single internal direction, $n=1$, we have $Q_4\notin \Z$, $\delta Q_4={1\over 2}$ and the above results reduce to 
\be
\begin{aligned}
\frac{1}{8\pi}\frac{\Lambda_{\rm ind}}{G_{\rm ind}}&= -4N \, \frac{\Gamma\big({5\over 2}\big)}{\pi^{{13\over2}}}\sum_{k}\frac{\e^{2i\pi  Q_4(2k+1)}}{\big|2k+1\big|^{5}}\, M_{\rm susy}^4+N{7i\over 2}\sum_{m_4} \delta^{(4)}(0)\, , \\
\frac{1}{8\pi G_{\rm ind}}&= N \, \frac{\Gamma\big({3\over 2}\big)}{6 \pi^{{9\over2}}}\sum_{k}\frac{\e^{2i\pi  Q_4(2k+1)}}{\big|2k+1\big|^{3}}\, M_{\rm susy}^2\, ,
\end{aligned}
\ee
where $M_{\rm susy}$ is the mass gap between bosonic and fermionic superpartners, i.e. the supersymmetry breaking scale,
\be
M_{\rm susy}\equiv M_{m_4}\Big(Q_4+{1\over 2}\Big)-M_{m_4}(Q_4)={1\over 2R_4}\, .
\ee


\section{Non-dynamical fields}
\label{aux}

Non-dynamical scalar fields are often introduced in classical theories for various purposes. For instance,  they can be considered to implement linear realizations of symmetries in tree-level actions. This is for example the case for supersymmetry, where they appear as auxiliary fields in off-shell definitions of supermultiplets. In the following, we show that at the quantum level, the alternative definitions of classical theories with or without non-dynamical fields  affect the ``unphysical'' infinite contributions proportional to $i\delta^{(4)}(0)$, but not the physical part of the cosmological term and the gravitational constant. We will specialize to the examples of non-dynamical fields appearing in chiral and vector multiplets of $\N_4=1$ supersymmetry. 


\subsection{\bm $F$-term}

The component fields of a chiral multiplet are a complex scalar $\Phi$, a Weyl fermion $\psi$ and an auxiliary complex scalar $F$. Considering a KK tower of such modes, the bosonic part of the action is~\cite{wb}
\be
S_{\Phi F}= \int \d^4x\sqrt{-g}\,  \sum_\m \Big[\!-g^{\mu\nu} \partial_\mu\Phi^*_\m\partial_\nu\Phi_{\vec m}+ F_\m^*F_\m + M_\m\big(\Phi_\m F_\m + \Phi_\m^*F_\m^*\big)\Big] .
\label{spf}
\ee
A convenient way to analyze this system is to disentangle the scalars by applying a field redefinition $\tau_\m= F_\m+M_\m \Phi_\m^*$, which yields 
\be
\begin{aligned}
S_{\Phi F}&\equiv S_{\Phi}+S_{\tau}\, , \\
\where \quad S_{\Phi}&=- \int \d^4x\sqrt{-g}\,  \sum_\m \Big[g^{\mu\nu} \partial_\mu\Phi^*_\m\partial_\nu\Phi_{\vec m}+M_\m^2\Phi^*\Phi\Big] ,\\
S_{\tau}&=\int \d^4x\sqrt{-g}\, \tau^*\tau\, .
\end{aligned}
\ee
Of course, decomposing $\Phi_{\vec m}$ into its real and imaginary parts,  $\Phi_{\vec m}\equiv (\phi_{\vec m}+i\tilde \phi_{\vec m})/\sqrt{2}$, $S_\Phi$ is nothing but the action of the real scalars $\phi_{\vec m}$, $\tilde \phi_{\vec m}$, as given in Eq.~\eqref{scalar}, while $S_\tau$ vanishes upon using the classical equations of motion
\be
\tau_\m\equiv F_\m+M_\m \Phi_\m^*=0\, , \qquad \tau_\m^*\equiv F_\m^*+M_\m \Phi_\m=0\, .
\ee
If the contributions to the effective gravity action  arising from  the integrated out complex scalars $\Phi_\m$ are simply twice those given in Eqs~\eqref{lphi} and~\eqref{gphi}, our goal is to see how the result is affected by integrating out  the auxiliary complex scalars $\tau_\m$. 

The trace of the stress-energy tensor derived from $S_\tau$ is independent of the metric, 
\be
g^{\mu\nu}T_{\mu\nu}^\tau\equiv {-2\over \sqrt{-g}}\, g^{\mu\nu}{\delta S_\tau\over \delta g^{\mu\nu}}=4\sum_\m \tau_\m^*\tau_\m\equiv T_\tau(x)\, ,
\ee
while the propagator in Minkowski space derived from $S_\tau$ is\footnote{The two-point function $\big\langle \big(F_\m(x)+M_\m \Phi_\m^*(x)\big)\big(F^*_{\m'}(y)+M_{\m'} \Phi_{\m'}(y)\big)\big\rangle$ can also be obtained from those derived from $S_{\Phi F}$, and which  are listed in  Ref.~\cite{wb}.}  
\be
\langle \tau_\m(x)\tau^*_{\m'}(y)\rangle = \delta_{\m,\m'}\, i\delta^{(4)}(x-y) \equiv   \delta_{\m,\m'}\big(\square -M_\m^2\big)\Delta(x-y)\, .
\ee
Thus, we obtain 
\be
\begin{aligned}
\langle T_{\tau}(0)\rangle&= 4\sum_\m  i\delta^{(4)}(0)\, , 
\\
\langle\widetilde T_\tau(x)\widetilde T_\tau(0)\rangle&=16\sum_\m \big(\square -M_\m^2\big)\Delta_\m (x) \,  \big(\square -M_\m^2\big)\Delta_\m (x)\, ,
\end{aligned}
\ee
and it is straightforward to derive the contributions to the cosmological and Newtonian constants,
\be
\begin{aligned}
\left.\frac{1}{8\pi}\frac{\Lambda_{\rm ind}}{G_{\rm ind}}\right|_{\tau} = -i\sum_\m \delta^{(4)}(0) \, ,\qquad \left.\frac{1}{8\pi G_{\rm ind}}\right|_{\tau} & =  2I_{3}({\vec Q})-\frac{16}{9}I_4({\vec Q})-\frac{32}{39}I_6({\vec Q})\\
&=0\,  ,
\end{aligned}
\label{taul}
\ee
where we have used the fact that all quantities $I_3({\vec Q})$, $I_4({\vec Q})$, $I_6({\vec Q})$ vanish. Hence, we obtain the announced result that 
the presence of the non-dynamical field $F$ in the definition of the classical theory affects only the number of infinite Dirac distributions at the origin, per KK mode $\m$. 
 

\subsection{\bm $D$-, $\cM$-, $\cN$-terms}

The second example we consider is that of the non-dynamical fields appearing in a vector multiplet of $\N_4=1$ supersymmetry. The component fields of this supermultiplet are a vector boson $A_\mu$, a real scalar $C$, 2 Weyl fermions $\lambda$, $\psi$, and three real auxiliary fields $D$, $\cM$, $\cN$.\footnote{In the massless case, $C$, $\psi$ and $\cM$, $\cN$ can be gauged away. However, in the massive case, the supermultiplet can be viewed as a massless vector multiplet in Wess-Zumino gauge, $(A_\mu,\lambda,D)$, and a massless chiral multiplet $(\Phi, \psi, F)$, where the vector boson ``eats'' one degree of freedom of $\Phi$, to let us with a massive vector field $A_\mu$, a degenerate scalar $C$, the Weyl fermions $\lambda$, $\psi$ and three real auxiliary scalars.}  For a tower of KK states with masses $M_\m$, the four-dimensional action restricted to the bosonic sector is~\cite{wb}
\be
\begin{aligned}
S_{\rm v.m.}=\int \d^4x&\sqrt{-g}\,  \sum_\m \Big[\!-{1\over 4}\, g^{\mu\nu}g^{\rho\sigma}F_{\m\mu\nu}F_{\m\rho\sigma}+{1\over 2}\, D_\m^2 \\
&+M_\m^2\Big( \!-{1\over 2}\,g^{\mu\nu}A_{\m\mu}A_{\m\nu}-{1\over 2}\, g^{\mu\nu}\partial_\mu C_\m\partial_\nu C_\m+C_\m D_\m+{1\over 2}\, \cM_\m^2+{1\over 2}\,\cN_\m^2 \Big)\Big].
\end{aligned}
\ee
This system can be divided into three pieces totally decoupled from one another.\footnote{We remind that they are not even coupled gravitationally, since the metric $g_{\mu\nu}$ is only a constant background and not a quantum field.}  The first subsystem amounts to the vector bosons $A^\mu_\m$, whose contributions to the effective gravity action are already given in Eqs~\eqref{al} and \eqref{gl}.  

Because $M_\m\neq 0$, we may redefine $\Phi_\m\equiv M_\m C_\m/\sqrt{2}$ and $F_\m\equiv D_\m/\sqrt{2}$, so that the dynamics of the second subsystem that comprises the scalars $C_\m,D_{\m}$ may be described by the action
\be
S_{CD}=\int \d^4x\sqrt{-g}\,  \sum_\m \Big[\!-g^{\mu\nu} \partial_\mu\Phi_\m\partial_\nu\Phi_{\vec m}+ F_\m F_\m + M_\m\big(\Phi_\m F_\m + \Phi_\m F_\m\big)\Big] .
\ee
The latter is nothing but $S_{\Phi F}$ given in Eq.~\eqref{spf}, but for real scalars $\Phi_\m$ and real non-dynamical fields $F_\m$. Therefore, the contributions of the fields $C_\m, D_{\m}$ to the induced Einstein action are those of a real scalar given in Eqs~\eqref{lphi} and \eqref{gphi}, plus those of the real part of $\tau$ given in Eq.~\eqref{taul}, namely
\be
\left.\frac{1}{8\pi}\frac{\Lambda_{\rm ind}}{G_{\rm ind}}\right|_{D} =  -{i\over 2}\sum_\m \delta^{(4)}(0) \, ,\qquad 
\left.\frac{1}{8\pi G_{\rm ind}}\right|_{D} =0\,  .
\ee

The last subsystem to be analyzed amounts to the fields $\cM_\m$ and $\cN_\m$. However, defining $\hat \tau_\m\equiv M_\m (\cM_\m+i\cN_\m)/\sqrt{2}$, we see that their  integration out is equivalent to that derived in the previous section, that is
\be
\left.\frac{1}{8\pi}\frac{\Lambda_{\rm ind}}{G_{\rm ind}}\right|_{\cM\cN} =  -i\sum_\m \delta^{(4)}(0) \, ,\qquad 
\left.\frac{1}{8\pi G_{\rm ind}}\right|_{\cM\cN} =0\,  .
\ee


\subsection{\bm Cancellation of the $i\delta^{(4)}(0)$-terms}

We have seen that the real non-dynamical  fields $X_\m$,\footnote{Such as $\Re \tau_\m/\sqrt{2}$, $\Im \tau_\m/\sqrt{2}$,  $D_\m$, $M_\m\cM_\m$, $M_\m\cN_\m$.} 
which have an action
\be
S_X=\int \d^4x\sqrt{-g}\,\sum_\m  {1\over 2}\,  X_\m X_\m\, , 
\ee
contribute to the effective gravitational constants  as
\be
\left.\frac{1}{8\pi}\frac{\Lambda_{\rm ind}}{G_{\rm ind}}\right|_{X} =  -{i\over 2}\sum_\m \delta^{(4)}(0) \, ,\qquad 
\left.\frac{1}{8\pi G_{\rm ind}}\right|_{X} =0\,  .
\ee
Because the degrees of freedom in Eq.~\eqref{spec} are identical to those corresponding to KK towers of $N$ massive vector multiplets, it may be natural\footnote{This is the case when the theory admits an underlying $\N_4=1$ supersymmetry, which can be spontaneously broken or not. However, as mentioned in Sect.~\ref{smod}, generic charge vectors of the KK towers break explicitly supersymmetry in four dimensions, and there is a priori no natural number of KK towers of non-dynamical fields.} that they are accompanied by $3N$ towers of non-dynamical real scalars. However, even in this case, this is not enough to cancel the infinite contribution
\be
N{7i\over 2}\sum_\m \delta^{(4)}(0)\, , 
\ee 
which appears in the induced cosmological term, Eq.~\eqref{resul}. Actually, $7N$ towers of non-dynamical fields are required, for which we obtain the fully finite result
\be
\begin{aligned}
\frac{1}{8\pi}\frac{\Lambda_{\rm ind}}{G_{\rm ind}}&=\sum_{u=1}^{N}I_0(\vec Q_u)-2\sum_{u=N+1}^{3N} I_0(\vec Q_u)+3\sum_{u=3N+1}^{4N}I_0(\vec Q_u)\, , 
\\ \frac{1}{8\pi G_{\rm ind}}&=-{1\over 3}\sum_{u=1}^{N}I_1(\vec Q_u)-{1\over 3}\sum_{u=N+1}^{3N} I_1(\vec Q_u)+\sum_{u=3N+1}^{4N}I_1(\vec Q_u)\, .
\end{aligned}
\label{resul2}
\ee


\section{Induced  higher-derivative terms}
\label{sak}

In this section, we reconsider the possibility of seeing Einstein gravity as a long-wavelength approximation of a more fundamental theory, when heavy matter fields are integrated out. We follow the point of view introduced by Sakharov~\cite{zah}, which provides an alternative derivation of the low energy effective action.  Because a lot is known about this approach~\cite{Schwinger,dewitt,Gilkey,Duff1,Duff2,Toms,BD,Avramidi1,Avramidi2,Avramidi3,
Visser,PT}, it will be straightforward to cross-check the results presented in Sec.~\ref{IEA}, and even extend them to include  higher-derivative terms. It turns out that integrating out towers of massive spin-0, spin-${1\over 2}$ and spin-1 quantum fields {\em does not} permit to cancel the UV divergencies occurring in 4-derivative terms. Therefore, we will introduce counterterms of the same form, namely classical kinetic terms for $g_{\mu\nu}$ with four derivatives.  In principle, there is no obstruction to adding such counterterms in the tree-level action when $g_{\mu\nu}$ is a classical field. However, we start this section by arguing that the metric should rather be treated as a quantum field. 


\subsection{Necessity to quantize gravity?}

Up to now,  the higher dimensional background metric $g_{MN}$, $M,N\in\{0,\dots,3+n\}$, has been treated as a pure classical field that influences the dynamics of the quantum scalars, fermions and vector bosons. Therefore, it is a matter of choice to take the components $g_{ij}\equiv R_i^2\delta_{ij}$ constant, the vector fields $g_{i\mu}\equiv 0$, and the four-dimensional metric $g_{\mu\nu}$ dependent on $x^\lambda$ only. 

However, the treatment of $g_{MN}(x^L)$ as a pure classical object implies some  drawbacks. First, there is no natural reason to impose any extremization principle on the effective action, with respect to the metric. In the setup presented in Sect.~\ref{IEA}, this means that Einstein equations  must  actually be imposed by hand on $g_{\mu\nu}(x^\lambda)$, as raised by Adler~\cite{Adler4}. Another issue is that if $g_{\mu\nu}$ is not quantized, the system of equations of motion of the classical metric and the quantum fields is not invariant under metric-dependent field redefinitions of the quantum matter degrees of freedom~\cite{Duff}. Therefore, we are led to quantize the metric. 

The approach we will follow from now on assumes that in addition to the matter fields, only $g_{\mu\nu}(x^\lambda)$ is to be quantized. In that case, our choice to take $g_{i\mu}\equiv 0$ and $g_{ij}\equiv R_i^2\delta_{ij}$ as constant parameters is still allowed. In practice, extremizing $S_{\rm eff}$ with respect to $g_{\mu\nu}(x^\lambda)$ yields Einstein theory, while no variational principle with respect to the radii is to be imposed. This very fact is fortunate, since  the $R_i$'s would otherwise behave as Lagrange multipliers that would  impose non suitable conditions.  The dissymmetry in the treatment of $g_{i\mu}$ and $g_{ij}$, as compared to the quantum field $g_{\mu\nu}$, and the fact that the latter depends only on the four coordinates $x^\lambda$, implies that the fundamental theory (i.e. before integrating out the matter fields) should now be considered as intrinsically four-dimensional.  It should be  regarded as ``already compactified'' and the infinite spectrum of spin-0, spin-${1\over 2}$ and spin-1 fields in four dimensions should only be formally seen as KK towers of states. 
In particular, there are neither KK states for the graviton, nor graviphotons and radion fields.


\subsection{Heat kernel expansion method}

From now on, our fundamental theory at tree-level is four-dimensional. Beside the gravitational sector, the quantum degrees of freedom comprise $N_\phi=N$ towers of scalars, $N_\psi=2N$ towers of Weyl fermions and $N_A=N$ towers of vector bosons of masses $M_\m(\vec Q_u)\equiv M_{u\m}$. As in Sect.~\ref{IEA}, we first assume that all charge vectors satisfy $\vec Q_u\notin \Z^n$. However, we will generalize our analysis at the end of this section to include the possibility of having  some $\vec Q_u\in\Z^n$.  The total classical action is 
\be
S_{\rm tree}=S_{\rm g}+\sum_{u=1}^{4N}S_u(\vec Q_u)\, ,
\label{str}
\ee
where $S_u(\vec Q_u)$ is of the form given in Eq.~(\ref{scalar}) for $u=1,\dots,N$, Eq.~(\ref{acfer}) for $u=N+1,\dots, 3N$, and Eq.~(\ref{acvb}) for $u=3N+1,\dots, 4N$. Moreover, $S_{\rm g}$ is a purely gravitational action to be determined later. Our goal is to derive the 1-PI effective action in the semiclassical limit for $g_{\mu\nu}$, which can be written as~\cite{BD}  
\be
S_{\rm eff}= S_{\rm tree}+W\, , 
\ee 
where $W$ is  the 1-loop contribution arising from the radiative corrections of the matter fields,
\be
W={i\over 2}\sum_u (-1)^F \sum_\m \ln \det \Big[(-D_u^2+M_{u\m}^2-i\varepsilon)\sigma^2\Big].
\label{w}
\ee 
Because the actions $S_u$ are quadratic in the integrated fields, $W$ is actually exact. In the above expression, $F=0$ for the bosonic towers $u$ and $F=1$ for the fermionic ones, while $D_u^2$ is the  kinetic operator, which contains 2-derivatives and depends on the metric. For instance, for a tower of scalars  $\phi_{u\m}$, we have $D_u^2\phi_{u\m}\equiv \square_g\phi_{u\m}\equiv {1\over \sqrt{-g}}\partial_\mu(\sqrt{-g}\, g^{\mu\nu}\partial_\nu\phi_{u\m})$. Finally, $\sigma$ is an arbitrary length  introduced for dimensional purpose. 

Let us notice that for any eigenvalue $K\sigma^2$ of the operator in square brackets in Eq.~(\ref{w}), which has a small negative imaginary part $-i\varepsilon$, one has 
\be
\ln (K\sigma^2) = -\int_{\rho^2}^{+\infty}{\d s\over s}\, \e^{-is K} -\ln \Big[i\e^{\gamma}\Big({\rho\over\sigma}\Big)^2\Big]+\O(\rho^2 K)\, , 
\ee
where $\rho$ is a length (an UV cutoff) and $\gamma$ is the Euler--Mascheroni constant. Using the  identity  $\ln \det \equiv \tr\! \ln$ and the above relation, we obtain 
\be
\begin{aligned}
W&=-{i\over 2}\sum_u (-1)^F\sum_\m \tr \int_{\rho^2}^{+\infty}{\d s\over s}\, \e^{-is(-D_u^2+M_{u\m}^2-i\varepsilon)}+W_1\, ,  \\
\where \quad  W_1&=\phantom{-}{i\over 2}\sum_u (-1)^F\sum_\m \tr \O\big[\rho^2(-D_u^2+M_{u\m}^2-i\varepsilon)\big]\underset{\rho\to 0}{\longrightarrow} 0\, . 
\end{aligned}
\ee
Notice that all contributions $\ln (i\e^{\gamma}(\rho/\sigma)^2)$  have cancelled out, due to the equal number of bosonic and fermionic towers of states. Because $W_1$ is irrelevant when $\rho\to 0$, we will omit it from now on.   To proceed, the key point is to use the heat kernel expansion~\cite{Schwinger,dewitt,BD} 
\be
-{i\over 2}\, \tr \e^{-is(-D_u^2+M_{u\m}^2-i\varepsilon)} =-{1\over 2} \int \d^4x \sqrt{-g} \;  {\e^{-is(M_{u\m}^2-i\varepsilon)}\over (4\pi s)^2} \, \sum_{\kappa=0}^{+\infty}\,(is)^\kappa a_{u\kappa}\, , 
\label{tra}
\ee
where $a_{u\kappa}$ depends on the spacetime geometry at $x$, and involves $2\kappa$ derivatives. For the lower values of $\kappa$, we have 
\be
\begin{aligned}
a_{u0}&=k_{u\Lambda}\, , \\
a_{u1}&=k_{u\cR}\, \cR\, , \\
\sqrt{-g}\, a_{u2}&=\sqrt{-g}\,\big(k_{u\cR^2}\,\cR^2+ k_{u\C^2}\, \C^2 \big)+\mbox{total derivative}\, ,
\end{aligned} \label{ff}
\ee
where $\cR$ is the Ricci scalar, and $\C^2\equiv \C_{\mu\nu\lambda\omega} \C^{\mu\nu\lambda\omega}$, with $\C_{\mu\nu\lambda\omega}$ the Weyl tensor. The coefficients appearing in the right hand sides of these equations depend on the spin of each  tower $u$, and  are given in Table~\ref{tabcoeff}~\cite{dewitt,Duff1,Duff2,BD}.
\begin{table}
\begin{centering}
\begin{tabular}{l|cccc}
 & $k_{u\Lambda}$ & $k_{u\cR}$ & $k_{u\cR^2}$ & $k_{u\C^2}$ \\
\hline
 spin-0 real scalars & $1$ & ${1\over 6}$ & ${1\over 72}$ & ${1\over 120}\esp $\\
spin-$\frac{1}{2}$ Weyl fermions & $2$ & $-{1\over 6}$ & $0$ & $-{1\over 40} \esp $\\
spin-1  massive vector bosons & $3$ & $-{1\over 2}$ & ${1\over 72}$ & $ {13\over 120}\esp\espD$\\
spin-1  massless vector boson & $2$ & $-{2\over 3}$ & $0$ & $ {1\over 10}\esp\espD$\\
\hline
\end{tabular}
\par\end{centering}
\caption{\footnotesize Values of the  coefficients $k_{u\Lambda},\, \, k_{u\cR},\, \, k_{u\cR^2}$ and $k_{u\C^2}$ appearing in Eq.(\ref{ff}) for each KK mode of the massive towers of states. The case of a massless vector boson is also included for reference. }
\label{tabcoeff}
\end{table}
Moreover, the total derivative term does not contribute to the equations of motions and can be ignored. Among other things, the latter contains a Gauss-Bonnet contribution, which is related to the Euler characteristic of the four-dimensional spacetime,
\be
\chi ={1\over 32\pi^2} \int \d^4x\sqrt{-g}\, \big(  \cR_{\mu\nu\rho\sigma} \cR^{\mu\nu\rho\sigma} - 4\cR_{\mu\nu}\cR^{\mu\nu} + \cR^2 \big)\, , 
\ee
where $\cR_{\mu\nu\rho\sigma}$ and  $\cR_{\mu\nu}$ are the Riemann and Ricci tensors. 

\begin{figure}
\begin{center}
\begin{tikzpicture}[scale=1.3]

\draw [->] (-1,0) -- (3.7,0) ;
\draw [->] (0,-3.7) -- (0,1) ;

\draw [ultra thick,red,>=latex,->] (0.5,0) -- (1.5,0) ;
\draw [ultra thick,red] (1.24,0) -- (3,0) ;
\draw [ultra thick,red] (3,0) arc (0:-90:3) ;
\draw [ultra thick,red] (0,-3) -- (0,-0.5) ;
\draw [ultra thick,red] (0,-0.5) arc (-90:0:0.5);

\draw (-25:3.3) node{$\red \cC$} ;
\draw (-45:0.7) node{$\cA$} ;
\draw (0.5,0.3) node{$\rho^2$} ;
\draw (3.4,0) node[above]{$\Re s$} ;
\draw (0,0.9) node[left]{$\Im s$} ;

\end{tikzpicture}
\end{center}
\caption{Contour integral used to derive Eq.~(\ref{inte}). }
\label{contour}
\end{figure}
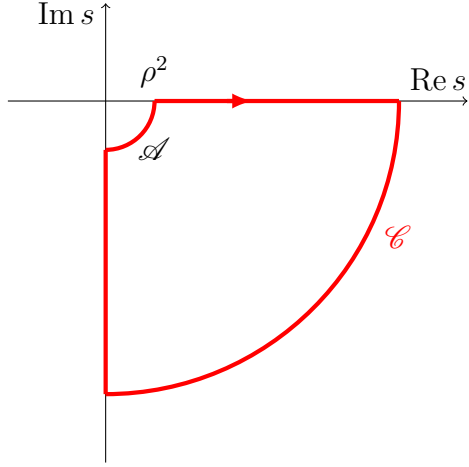

The 1-loop contribution $W$ to the effective action can be decomposed as 
\be
\begin{aligned}
&W=\int \d^4x \sqrt{-g} \,\sum_{\kappa=0}^{+\infty}\L_\kappa\, , \\
\where\quad &\L_\kappa(x)={1\over 32 \pi^2} \sum_u (-1)^F a_{u\kappa}(x)\sum_\m \int_{\rho^2}^{+\infty}i\d s \, f_{u\kappa \m}(is)\, , \\
& f_{u\kappa \m}(is)=(is)^{\kappa-3}\, \e^{-is(M_{u\m}^2-i\varepsilon)}  \, .
\end{aligned}
\label{Lags}
\ee
In the Lagrangian density $\L_\kappa$, the integral can be expressed as 
\be
\label{inte}
\int_{\rho^2}^{+\infty}i\d s\, f_{u\kappa \m}(is)=\int_{\rho^2}^{+\infty}\d t\, f_{u\kappa \m}(t)+\int_\cA i\d s\, f_{u\kappa \m}(is)\, , 
\ee
where we have applied the change of variable $t=is$, and $\cA$ is the small circular arc of radius $\rho^2$ shown in Fig.~\ref{contour}. 
The above identity is proven by noticing that $f_{u\kappa \m}$ has no pole inside the contour $\cC$ of Fig.~\ref{contour}, implying the integral over $\cC$ to vanish.  The rest of the discussion depends on the degree of singularity of the functions $f_{u\kappa \m}$:
\begin{itemize}
\item For $\kappa\ge 3$, $f_{u\kappa \m}$ has no pole and the introduction of $\rho$ was actually not necessary. Taking the limit $\rho\to 0$ in the r.h.s. of Eq.~(\ref{inte}), only the Wick-rotated integral over $t$ survives, and  we find the finite result
\be
\forall \kappa\ge 3:\quad \L_\kappa = {\Gamma(\kappa-2)\over 32\pi^2}\sum_u (-1)^F \sum_\m {1\over M_{u\m}^{2(\kappa-2)}}\, a_{u\kappa}\, .
\ee
\item For $\kappa=0,1$, we invert the discrete sums over $\m$ and the integrals, in order to use the Poisson summation  formula~(\ref{pr}). The latter yields the expression 
\be
\sum_\m f_{u\kappa \m}(t)={\pi^{n\over 2}\over t^{3+{n\over 2}-\kappa}}\Big(\prod_i R_i\Big)\sum_\vell \e^{-{\pi^2\over t} \sum_j(\ell_jR_j)^2}\e^{2i\pi \vec Q_u\cdot\vell}\, ,
\label{po}
\ee
where all poles arise from the term $\vell =\vec 0$. However, summing over the  towers of states, the coefficients of the contributions $\vell=\vec 0$ vanish,
\be
\begin{aligned}
&\sum_u (-1)^F k_{u\Lambda}= N_\phi-2N_\psi+3N_A=0\, , \\
&\sum_u (-1)^F k_{u\cR}= {1\over 6}N_\phi-\Big(\!-{1\over 6}\Big)N_\psi-{1\over 2}N_A=0\, ,
\end{aligned} \label{d1}
\ee 
and it is safe to take the limit $\rho\to 0$. The integrals that survive are the Wick-rotated ones over $t$, and they can be computed term by term using the change of variable $l:=\pi^2\sum_j(\ell_j R_j)^2/t$. The final results for the Lagrangian densities are 
\be
\begin{aligned}
\L_0&\equiv-\frac{1}{8\pi}\frac{\Lambda_{\rm ind}}{G_{\rm ind}}\;= {\Gamma(2+{n\over 2})\over 32\pi^{6+{n\over 2}}}\Big(\prod_{i} R_i\Big)\sum_u (-1)^Fk_{u\Lambda}\sum_{\vell\neq \vec 0}\frac{\e^{2i\pi \vec Q_u\cdot\vell}}{\big(\sum_j\ell_j^2R_j^2\big)^{2+{n\over2}}}\, ,\\
\L_1&\equiv \frac{1}{8\pi G_{\rm ind}}\,  {\cR\over 2}={\Gamma(1+{n\over 2})\over 16\pi^{4+{n\over 2}}}\Big(\prod_{i} R_i\Big)\sum_u (-1)^Fk_{u\cR}\sum_{\vell\neq \vec 0}\frac{\e^{2i\pi \vec Q_u\cdot\vell}}{\big(\sum_j\ell_j^2R_j^2\big)^{1+{n\over2}}}\, {\cR\over 2}\, ,
\end{aligned}
\ee
which are in perfect agreement with the induced cosmological and gravitational  constants of Eq.~(\ref{resul2}). 
\item For $\kappa=2$, Eq.~(\ref{po}) applies  but the dangerous contributions $\vell=\vec 0$ do not vanish, since 
\be
\begin{aligned}
&\sum_u (-1)^F k_{u\cR^2}= {1\over 72}N_\phi-0\cdot N_\psi+{1\over 72}N_A={N\over 36}\, ,\\
&\sum_u (-1)^F k_{u\C^2}\,=  {1\over 120}N_\phi-\Big(\!-{1\over 40}\Big)N_\psi+{13\over 120}N_A={N\over 6}\, .
\end{aligned}
\label{nan}
\ee 
As a result, the 1-loop Lagrangian density  $\L_2$ given in Eq.~(\ref{Lags}) is UV divergent, as $\rho\to 0$. To make sense, it must be combined, with tree-level $\C^2$- and $\cR^2$-terms with infinite bare couplings $f^2_{0\rm B}$, $f^2_{2\rm B}$. Hence, the purely gravitational tree-level action is defined as
\be
S_{\rm g}=\int \d^4x\sqrt{-g}\left[{\cR^2\over 6f^2_{0\rm B}}-{\C^2\over 2f^2_{2\rm B}}\right]\! . 
\ee
Notice that had we chosen other values of  $N_\phi$, $N_\psi$, $N_A$, the UV divergence of $\L_2$ would not have vanished, as can be seen from Eq.~(\ref{nan}).   
\end{itemize}
Collecting all of the above results, the 1-PI effective action reads   
\be
S_{\rm eff}=\int \d^4x \sqrt{-g} \left[{1\over 8\pi G_{\rm ind}}\Big({\cR\over 2}-\Lambda_{\rm ind}\Big)+{\cR^2\over 6f^2_{0}}-{\C^2\over 2f^2_{2}}+\sum_{\kappa=3}^{+\infty}\L_\kappa+\sum_{u=1}^{4N}S_u(\vec Q_u)\right]\!,  
\ee
where $f^2_0$, $f_2^2$ are finite, renormalized couplings to be determined by measurements. On the other hand, the cosmological and Newtonian constants, as well as all couplings of $2\kappa$-derivative terms with  $\kappa\ge 3$, are calculable in terms of the $n$ radii $R_i$ and the $4N$ charge vectors $\vec Q_u$. 

Notice that at this stage, we have treated the gravitational degrees of freedom semiclassically. When $f_0^2$, $f_2^2$ are positive, the latter amount to the massless graviton, a real scalar field and a spin-2 ghost graviton.\footnote{Even if it is not fully settled in the literature,  it is sometimes believed that the ghost degrees of freedom don't  appear as external states of the $S$-matrix, and that they don't spoil unitarity~\cite{Salvio, AG, Don2}.}  
In the pure quadratic gravity case~\cite{Stelle1,Stelle2}, i.e. when the kinetic terms are restricted to be of the form $\cR^2$ and $\C^2$ only, and when matter fields are not present, the theory is scale invariant. Therefore, dimensionful parameters such as a cosmological constant and a reduced Planck mass $M^2_{\rm P}\equiv 1/(8\pi G)$ cannot be generated at the quantum level, at any order of perturbation theory. For instance, this can be checked at the 1-loop level by considering the associated renormalization group equations~\cite{Avramidi1,CP,Salvio2,Salvio},
\be
\begin{aligned}
{\d M^2_{\rm P}\over \d\tau}&=\left({2\over 3}\, f_0^2-{5\over 3}\, {f_2^4\over f_0^2}\right)\! M_{\rm P}^2\, , \\
{\d (\Lambda M^2_{\rm P})\over \d\tau}&={5f_2^4+f_0^4\over 8}\, M_{\rm P}^4+(5f_2^2+f_0^2)\Lambda M^2_{\rm P}\, , 
\end{aligned}
\ee
where $\tau\equiv \ln(\mu/\mu_0)/(4\pi)^2$, $\mu $ is the energy scale in $\overline{\rm MS}$ scheme, and $\mu_0$ is an arbitrary fixed energy. From the above expressions, it is clear that $\Lambda M_{\rm P}^2=0$, $M_{\rm P}^2=0$ is a fixed point of the renormalization group. However, in the theory considered throughout the present work, additional diagrams at 1-loop arise from a specific matter spectrum and yield finite, non-running constants $\Lambda=\Lambda_{\rm ind}$ and $M_{\rm P}^2=1/(8\pi G_{\rm ind})$.   The key point is that at 1-loop, Feynman diagrams associated with different sectors of the theory add linearly. Of course, one has to raise the question of whether the induced Einstein terms remain predictable at higher loop or not. As noticed before, higher order diagrams  involving only gravitational propagators cannot yield corrections to $\Lambda$ and $M_{\rm P}^2$. Nevertheless, those mixing free KK matter and gravitational states should be analyzed with scrutiny.  

Before closing this section, we would like to generalize  our results to include the case where some towers of states are characterized by  charge vectors $\vec Q_u\in\Z^n$, which yield low lying massless states. Notice that this is not possible  in Adler's approach, where all matter fields are integrated out. However,  in the 1-PI effective action, radiative corrections associated with  massless states can be taken into account. In Eq.~(\ref{tra}), we have used the fact that when the charges are non-integers, the heat kernel coefficients $a_{u\kappa}$ are independent of the modes $\vec m$. Hence, as can be seen from Table~\ref{tabcoeff}, our derivations still hold for towers of real scalars and Weyl fermions when $\vec Q_u\in\Z^n$. On the contrary, the heat kernel coefficients for massless and massive vector bosons differ. However, combining a massless vector and a massless real scalar field, the total coefficients match with those of a massive gauge boson (see Table~\ref{tabcoeff}). Of course, this is not a coincidence since every massive KK vector boson arises by absorbing one degenerate real scalar degree of freedom. Therefore, in order to accommodate a tower of vector bosons with $\vec Q_u\in \Z^n$, we add from the outset a degree of freedom $\phi_u$,  with classical action   
\be
S_{u}^0=-\int \d^4x\sqrt{-g}\,  \frac{1}{2}\,g^{\mu\nu}\partial_\mu\phi_u\partial_\nu\phi_u\, .
\ee


\section{Conclusion}
\label{conc}

In this work, we  have considered a four-dimensional gravity action based on four-derivative kinetic terms coupled to infinite towers of free massive  scalar, fermionic and vector fields. In particular, no cosmological and Einstein-Hilbert terms are present at tree-level. At the quantum level, though, we found that predictable induced cosmological constant $\Lambda_{\rm ind}$ and Newton constant $G_{\rm ind}$  are generated by integrating out the infinite massive states. $\Lambda_{\rm ind}$ and $G_{\rm ind}$ are expressed in terms of the radii and charges that specify the mass spectrum. Hence,  the standard Einstein gravity  is recovered at large distances. This is achieved only for a specific number of towers of scalars, fermions and vectors. We have used two methods for our calculations with identical results, namely Adler's point of view~\cite{Adler1,Adler2,Adler3,Adler4,Adler5} and heat kernel methods~\cite{Schwinger,dewitt,Gilkey,Avramidi3,BD,PT}. The bottom line of our approach is that the finite and not-running values of $\Lambda_{\rm ind}$ and $G_{\rm ind}$ are only induced  by towers of matter fields, and that there are no KK modes associated with the graviton. In other words, spacetime is ``already compactified'',  which is reminiscent of F-theory~\cite{Vafa}, where the fields in ten dimensions, including the graviton, are not accompanied with KK towers of states arising from the underlying twelve-dimensional spacetime. 

As long as the gravitation degrees of freedom are treated semiclassically, our effective action is exact in the sense that it is obtained by computing Gaussian path integrals of free massive fields. Our results for $\Lambda_{\rm ind}$ and $G_{\rm ind}$ remain valid once 1-loop effects in the gravitational sector are included. However, higher-loop corrections are beyond the scope of the present work.

A natural question to ask is whether our scenario may be generalized to account for the Standard Model (SM). If the SM degrees of freedom were introduced without towers of states, the induced parameters $\Lambda_{\rm ind}$, $G_{\rm ind}$ would no longer be invariant under the renormalization group~~\cite{Avramidi1,CP,Salvio2,Salvio}. Hence, the SM fields should be seen as being part of the towers of modes considered in the present work. The Higgs scalar, the $W^{\pm}, Z^0$ bosons and the fermions may correspond to low lying states, for choices of vectors $\vec Q_u\notin \Z^n$ that reproduce their classical masses (see Eq.~(\ref{mm})). On the contrary, for the massless photon and gluons, one should take $\vec Q_u\in \Z^n$. Of course,  the low lying SM fields (massive or massless) should not be integrated out, and the effective action should be 1-PI. In order for the induced cosmological and Newton constants to be calculable, we may supply the SM towers with other towers in a ``hidden sector'', such that the rule of thumb given in Eq.~(\ref{spec}) is satisfied. Moreover, an extra real scalar should be included in the hidden sector for every  tower of vectors fields such that  $\vec Q_u\in \Z^n$. However, let us remind that the analysis of the present paper has been derived for free matter fields. Hence, it is fair to stress that in order to accommodate the SM, one should extend our derivations to include interactions and therefore quantum corrections of matter beyond 1-loop. 


 \section*{Acknowledgements}
 
We are grateful to Pierre Fayet and Guillaume Bossard for lively discussions and useful inputs during the realization of this work. 
The work of H.P. is partially supported by the Royal-Society/CNRS International Cost Share Award IE160590.


\section*{Appendix}
\renewcommand{\theequation}{A.\arabic{equation}}
\renewcommand{\thesection}{A}
\setcounter{equation}{0}

In this appendix, we detail the prescriptions we follow to evaluate the discrete KK sums and integrals over momenta or spacetime coordinates that are encountered in the computations of the induced cosmological and gravity constants. A cutoff $\Lambda_{\rm max}$ in momentum can be introduced from the beginning of the computations. However, because our final results are finite in the limit $\Lambda_{\rm max}\to +\infty$, for the sake of notational simplicity, we will be cavalier by writing all  expressions with  $\Lambda_{\rm max}$ infinite.  
In all computations, our first step consists  in applying Wick rotations $x^0=-ix^0_{\rm E}$, $k_0=i k_{{\rm E}0}$ to convert  all Lorentzian scalar products into Euclidean ones. Then, we introduce Schwinger parameters to every Feynman propagator:
 \be
-i\int\frac{\d^4k}{(2\pi)^4}\, \frac{\e^{ik\cdot x}}{k^2+M_\m^2-i\varepsilon}=\int\frac{\d^4k_{\rm E}}{(2\pi)^4}\, \frac{\e^{ik_{\rm E}\cdot x_{\rm E}}}{k_{\rm E}^2+M_\m^2}=\int\frac{\d^4k_{\rm E}}{(2\pi)^4} \int_0^{+\infty}\!\!\!\!\!\!\!\d t\, \e^{-k_{\rm E}^2t+ik_{\rm E}\cdot x_{\rm E}}\e^{-M_\m^2t}\, .
\label{ps}
\ee
In the above equalities, it is understood that the initial UV cutoff in momentum translates into a cutoff as $t\to 0$ in the Schwinger integral. 

As an example, let us consider the second term of the correlator given in Eq.~\eqref{tphi}, with normalization as given in Eq.~\eqref{adler_form} for describing a contribution to the cosmological constant arising from a KK tower of real scalar fields, 
 \be
I_0({\vec Q})={1\over 4} \sum_\m M_\m^2\, \Delta_\m (0)\, .
\ee
Applying Eq.~(\ref{ps}), we then invert the integrals over the Euclidean momentum and $t$, and compute the Gaussian  integral over $k_{\rm E}$.   Next, we invert the discrete sum over the KK modes $\m$ and the integral over $t$. Applying the  following Poisson summation formula on~$\m$,
\be
\begin{aligned}
\sum_\m t\sum_i\Big({m_i+Q_i\over R_i}\Big)^2\e^{-t\sum_j\!\big({m_j+Q_j\over R_j}\big)^2} \!\!=&\; {\pi^{n\over2}\over 2\, t^{n\over 2}}\Big(\prod_{i} R_i\Big)\!\sum_{\vell}\e^{-{\pi^2\over t} \sum_j(\ell_jR_j)^2}\e^{2i\pi \vec Q\cdot\vell}\\
&\qquad\;\, \qquad \qquad \times \Big( n - {2\pi^2\over t}\sum_k(\ell_kR_k)^2\Big) ,
\end{aligned}
\ee
which can be derived from the one-dimensional identity 
\be
\sum_m \e^{-t\big({m+Q\over R}\big)^2} ={\pi^{1\over2}\over t^{1\over 2}}\, R \sum_{\ell}\e^{-{\pi^2\over t} (\ell R)^2}\e^{2i\pi Q\ell}\, ,
\label{pr}
\ee
where $m,\ell\in\Z$,  we find 
\be
I_0({\vec Q})= {(\prod_i R_i)\over 2^7\pi^{2-{n\over 2}}}\int_0^{+\infty}\!\!\!{\d t\over t^{3+{n\over 2}}}\sum_\vell \e^{-{\pi^2\over t} \sum_j(\ell_jR_j)^2}\e^{2i\pi \vec Q\cdot\vell}\,\Big( n - {2\pi^2\over t}\sum_k(\ell_kR_k)^2\Big).
\ee
Notice that the divergence is now fully concentrated in the $\vell=\vec 0$ term. Keeping in mind that this pathological contribution will cancel out between different KK towers, we proceed by integrating term by term over $l:=\pi^2\sum_j(\ell_j R_j)^2/t$, which  is allowed for $\vell\neq \vec 0$ and only formal for $\vell=\vec 0$. The result is 
\be
I_0({\vec Q})=  {{\pazocal I}(n)\over 2^7\pi^{6+{n\over 2}}}\,\Big(\prod_{i}R_i\Big)\sum_{\vell}\frac{\e^{2i\pi \vec Q\cdot\vell}}{\big(\sum_j\ell_j^2R_j^2\big)^{2+{n\over2}}}\, ,
\label{i0}
\ee
where 
\be
{\pazocal I}(n)=\int_0^{+\infty}\!\!\!\d l\, l^{1+{n\over 2}} \, \e^{-l}\,  (n-2l)=-4\, \Gamma\Big(2+{n\over 2}\Big) . 
\ee

As another example, let us compute $I_1({\vec Q})$ given in Eq.~\eqref{integrals}, which is part of the gravitational constant arising by integrating out a KK tower of real scalar fields. Using Eq.~\eqref{ps} and moving the Schwinger integrals to apply last, we have
\be
\begin{aligned}
I_1({\vec Q})& = -\frac{1}{24\, (2\pi)^8}\int_0^{+\infty}\!\!\!\!\!\!\!\d t_1\d t_2\int\d^4x_{\rm E}\,x_{\rm E}^2\sum_{\vec m}\e^{-(t_1+t_2)M_{\vec m}^2}\, {\pazocal J}(t_1,t_2,x_{\rm E})\, , \\
\where \quad {\pazocal J}(t_1,t_2,x_{\rm E})&=\int \d^4p_{\rm E}\, \d^4k_{\rm E}\, (p_{\rm E}\cdot k_{\rm E})^2\e^{-p_{\rm E}^2 t_1+ip_{\rm E}\cdot x_{\rm E}} \e^{-k_{\rm E}^2 t_2+ik_{\rm E}\cdot x_{\rm E}}\, .
\end{aligned}
\ee
Performing the Gaussian integrals over $p_{\rm E}$ and $k_{\rm E}$, we obtain 
\be
\begin{aligned}
{\pazocal J}(t_1,t_2,x_{\rm E})= {\pi^4\over (t_1t_2)^3}\!\left[1-{x_{\rm E}^2\over 8}\Big({1\over t_1}+{1\over t_2}\Big)+{x_{\rm E}^2\over 16}\, {1\over t_1t_2}\right] \!\e^{-{x_{\rm E}^2\over 4}\left({1\over t_1}+{1\over t_2}\right)}\, .
\end{aligned}
\ee
Because the integral over $x_{\rm E}$ is now Gaussian, it is straightforward to derive 
\be
{\pazocal K}(t_1,t_2)\equiv\int \d^4x_{\rm E}\, x_{\rm E}^2 \, {\pazocal J}(t_1,t_2,x_{\rm E})= -{2^6\pi^6\over (t_1+t_2)^3}\left[1-24\, {t_1t_2\over (t_1+t_2)^2}\right].
\ee
The next step is to apply the Poisson summation formula~\eqref{pr} on the KK sum, which yields
\be
I_1({\vec Q}) = -\frac{\pi^{{n\over 2}}}{24\, (2\pi)^8}\,\Big(\prod_i R_i\Big)\int_0^{+\infty}{\d t_1\d t_2\over (t_1+t_2)^{n\over 2}}\sum_{\vell}\e^{-{\pi^2\over t_1+t_2}\sum_j(\ell_j R_j)^2}\e^{2i\pi \vec Q\cdot\vell}\, {\pazocal K}(t_1,t_2)\, .
\ee
As before, the term $\vell=\vec 0$ contains whole of the UV divergence.  This follows from the fact that all other contributions, $\vell \neq 0$, can be integrated term by term, due to the exponential suppression as $t_1,t_2\to 0$. Note that integrating term by term before Poisson summation is not allowed, since the Schwinger integrals for each individual mode $\m$ is divergent at $t_1$ or $t_2= 0$.   
Because the $\vell=\vec 0$ contributions arising from different KK towers will cancel out, we proceed by integrating term by term over $\alpha_i =t_i/(\pi^2\sum_j(\ell_j R_j)^2)$, $i=1,2$, which leads to 
\be
\begin{aligned}
I_1({\vec Q}) &= \frac{\big(\prod_i R_i\big)}{3\cdot 2^5\, \pi^{4+{n\over 2}}}\sum_{\vell}\frac{\e^{2i\pi \vec Q\cdot\vell}}{\big(\sum_j\ell_j^2R_j^2\big)^{1+{n\over2}}}\, {\pazocal L}(n)\, , \\
\where \quad {\pazocal L}(n)&= \int_0^{+\infty}\!\!\!\!{\d\alpha_1\d \alpha_2\over (\alpha_1+\alpha_2)^{3+{n\over 2}}}\e^{-{1\over \alpha_1+\alpha_2}}\left[1-24\, {\alpha_1\alpha_2\over (\alpha_1+\alpha_2)^2}\right].
\end{aligned}
\ee
Defining $l=1/(\alpha_1+\alpha_2)$, $u=\alpha_1/(\alpha_1+\alpha_2)$, the last integral is found to be 
\be
{\pazocal L}(n)= \int_0^{+\infty}\!\!\!\d l \, l^{n\over 2}\e^{-l}\int_0^1 du \big[1-24 u(1-u)\big]\!=-3\, \Gamma\Big(1+{n\over 2}\Big).
\ee

For completeness, we mention that all other expressions $I_2({\vec Q}),\dots,I_6({\vec Q})$ can be computed in a similar way, except that $I_3({\vec Q})$ necessitates the use of the more involved Poisson summation formula
\begin{align}
\sum_\m t\sum_i\Big({m_i+Q_i\over R_i}\Big)^2&t\sum_j\Big({m_j+Q_j\over R_j}\Big)^2\e^{-t\sum_k\!\big({m_k+Q_k\over R_k}\big)^2}\nonumber \\
=&\; {\pi^{n\over2}\over 4\, t^{n\over 2}}\Big(\prod_{i} R_i\Big)\!\sum_{\vell}\e^{-{\pi^2\over t} \sum_j(\ell_jR_j)^2}\e^{2i\pi \vec Q\cdot\vell} \\
&  \times \bigg[ n(n+2) -4(n+2) {\pi^2\over t}\sum_k(\ell_kR_k)^2+4{\pi^4\over t^2}\Big(\sum_k(\ell_kR_k)^2\Big)^2\bigg].\nonumber 
\end{align}


\bibliographystyle{unsrt}


\end{document}